\documentclass[showpacs,amsmath,pre]{revtex4}
\usepackage{graphicx}
\usepackage{amsmath}

\usepackage{epsfig}
\usepackage{graphics}
\usepackage{latexsym}
\usepackage{amsfonts}
\usepackage{amssymb}
   
\usepackage{plain}    
  
\begin{document}

\title{Helium condensation in aerogel: avalanches and disorder-induced phase transition}

\author{F. Detcheverry, E. Kierlik, M. L. Rosinberg, and G. Tarjus}
\affiliation{Laboratoire de Physique Th\'eorique de la Mati\`ere Condens\'ee, Universit\'e Pierre et
Marie Curie, 4 place Jussieu, 75252 Paris Cedex 05, France}
  
\date{\today}

\begin{abstract}
We present a detailed numerical study of the elementary condensation events (avalanches) associated to the 
adsorption  of  $^4$He in silica aerogels. We use a coarse-grained lattice-gas description 
and determine the nonequilibrium behavior of the adsorbed gas within a local mean-field analysis, neglecting
thermal fluctuations and activated processes.
We investigate the statistical properties of the avalanches, such as their number, size and shape
along the adsorption isotherms as a function of gel porosity, temperature, and chemical potential. Our calculations predict the existence of a line of critical points in the temperature-porosity diagram where  the 
avalanche size distribution displays a power-law behavior and the adsorption isotherms have a universal scaling form.
The estimated critical exponents seem compatible with those of  the field-driven Random Field Ising Model at zero temperature.

\end{abstract}
\maketitle

\section{Introduction}
  
This work is part of  a series of papers devoted to the study of gas adsorption and capillary condensation in silica aerogels\cite{DKRT2003,DKRT2004}.  Our  goal is to provide a theoretical interpretation of   $^4$He adsorption experiments that have been performed over the last 15 years in several low-temperature physics groups\cite{WC1990,TYC1999,BH2003,LGPW2004}.  More generally, we wish to  propose a theoretical framework for describing the {\it nonequilibrium} (hysteretic) behavior of fluids in disordered porous solids\cite{KMRST2001}. It is worth noting that the experimental studies of  helium in aerogel were originally aimed at understanding the effect of quenched random impurities on {\it equilibrium} critical phenomena and phase separation.  In particular, the study of the liquid-vapor transition of $^4$He followed from the suggestion that the critical behavior of mixtures or pure fluids confined in porous media or in contact with a gel network could be interpreted in terms of the random field Ising model (RFIM)\cite{BG1983}. In this perspective, the use of aerogel was motivated by its very tenuous structure (which minimizes the effects of confinement) and by the fact that the porosity (and thus the amount of disorder) can be varied in a controlled way.  However, in spite of the early claim that mesoscopic equilibrium phase separation of $^4$He could be observed in a $95\%$ porosity aerogel\cite{WC1990}, leading to an estimation of the order parameter critical exponent $\beta$, it now seems more and more questionable that true equilibrium behavior is observed in experiments. In particular, recent measurements of sorption isotherms using a high-precision capacitive technique\cite{BH2003} show that hysteresis between filling and emptying persists to temperatures very close to the critical temperature of pure helium ($T_c=5.195K)$,  and gradual adsorption is always observed above $T=4.88K$, instead of the sharp vertical step expected for two-phase coexistence. Similar observations have been made by  another group\cite{LGPW2004}, using  a low-frequency mechanical pendulum technique and, more recently, coupling pressure and optical measurements. These results are reminiscent of the behavior of fluids confined in dense porous glasses which is usually interpreted in terms of capillary condensation.  Although this moves away the exciting perspective of studying the putative equilibrium vapor-liquid critical point of $^4$He inside aerogel, the remarkable changes in the hysteretic behavior of the adsorbed fluid with temperature and porosity remain to be explained.  In particular, the hysteresis loops in very light aerogels have a  rectangular (instead of triangular) shape  at very low temperature (for instance,  at $T=2.34K$ in  $98\%$  porosity aerogel\cite{TYC1999}) with a very steep adsorption branch that suggests the presence of a genuine, but {\it nonequilibrium}, first-order phase transition. This is in agreement with our theoretical description\cite{DKRT2003,DKRT2004} which predicts the change from a continuous to a discontinuous adsorption isotherm as one decreases the temperature (at constant porosity) or increases the porosity (at constant temperature), the jump in the fluid density corresponding to a macroscopic avalanche in the system. (Note that we only focus here on adsorption. The behavior on desorption is also interesting but is related to other physical processes as discussed in Refs.\cite{DKRT2003,DKRT2004}.)  This scenario, which resembles the one predicted for the field-driven RFIM at zero temperature\cite{S1993,PDS1996} and observed in some disordered magnetic systems\cite{B2000,MVMA2003}, suggests the existence of a line of disorder-induced critical points in the temperature-porosity diagram with an associated scaling behavior.
In the present work we further elaborate on this issue by presenting new results concerning  the avalanche properties  as a function of temperature and porosity and by studying the scaling behavior of the adsorption isotherms in the vicinity of the  nonequilibrium critical points. We are also interested in characterizing the avalanches geometrically and studying their possible relation with the aerogel structure.

The paper is arranged as follows. In section 2 we review the model and the theory. In section 3 we present the results for the statistical properties of avalanches. In section 4 we discuss the scaling behavior of the adsorption isotherms and present  the phase diagram of the model in the temperature-porosity plane. We conclude, in section 5, with a discussion of  the relevance of our results to experiments.

\section{Model and theory}

Since both the model and the theory were described in detail in ealier work (see in particular  Ref.\cite{DKRT2003}), we will restrict the presentation  to some key aspects that are relevant to our present goal. 

The gel-fluid system is modelled by a  {\it coarse-grained} lattice-gas where each  cell  of the size of a silica particle (typically a few nanometers\cite{DKRT2003}) is occupied either by the solid or the fluid. The distribution of silica particles on the lattice is obtained by a diffusion-limited cluster-cluster aggregation (DLCA) algorithm\cite{M1983}, so as to model the intricate microstructure of actual base-catalized silica aerogels. On the other hand, fluid particles can equilibrate with a reservoir that fixes their chemical potential $\mu $ and the temperature $T$. Multiple occupancy of a site is forbidden  and only nearest-neighbors (nn) attractive interactions are considered. For reasons that are explained in Ref.\cite{DKRT2004}, we use a body-centered cubic (bcc) lattice and periodic boundary conditions ($N=2L^3$ where $L$ is the linear size).

For a given configuration of the solid matrix, fluid configurations are obtained by minimizing a  grand-potential functional in the local mean field approximation\cite{KMRST2001},
\begin{align}
\Omega[\{\rho_i\}]&=k_BT \sum_{i} [\rho_i\ln \rho_i + (\eta_i-\rho_i)\ln (\eta_i-\rho_i)]
 -w_{ff}  \sum_{<ij>} \rho_i\rho_j   -w_{sf}  \sum_{<ij>} [ \rho_i(1-\eta_j) + \rho_j(1-\eta_i)]-\mu \sum_{i} \rho_i
\end{align}
where $\rho_i$ ($i=1...N$) is the thermally averaged fluid density at site $i$ and $\eta_i=0,1$ is the quenched occupancy variable for the gel particles ($\eta_i=0$ if the site is occupied by the gel and $1$ otherwise). $w_{ff}$ and  $w_{sf}$  denote the  fluid-fluid  and   solid-fluid interactions, respectively, and  the double summations run over all distinct pairs of nn sites. The gel porosity is given by $\phi=(1/N)\sum_i\eta_i$ and the ratio  $y=w_{sf}/w_{ff}$ controls the wettability of the solid surface.

$\Omega[\{\rho_i\}]$  thus defines a grand-potential landscape  whose extrema are obtained by solving the corresponding Euler-Lagrange equations $\delta \Omega/\delta \rho_i=0$. At fixed $T$ and $\mu$, and for a given realization of the solid, this yields a set  of $N$ coupled nonlinear equations

\begin{align}
\rho_i=\frac{\eta_i}{1+e^{-\beta (\mu+w_{ff}\sum_{j/i}[\rho_j+y(1-\eta_j)]) }}\ ,
\end{align}
where $\beta=1/(k_BT)$ and the sum runs over the nearest neighbors of
site $i$. 
A crucial feature of these equations is that they may have a large number of solutions at low temperature (and in a certain range of $\mu$) as a consequence of the randomness of the solid matrix\cite{note2}. We believe that this feature  is the key for understanding the physics of adsorption in disordered porous solids.
Especially important are the local {\it minima} of $\Omega[\{\rho_i\}]$ which are obtained by solving  the set of Eqs. (2) iteratively\cite{note3} and which correspond to metastable states (to be distinguished from the global minimum of $\Omega[\{\rho_i\}]$ that defines the true equilibrium state). Note that despite its mean-field character (by which thermal fluctuations are only crudely described), the present approach fully accounts for the disorder-induced fluctuations. In contrast with the standard global mean-field approximation, our treatment still allows to describe strongly inhomogenous configurations and to predict non-classical critical behavior. 

Computing the metastable states is however not sufficient for describing the nonequilibrium behavior of the system, and one also needs to specify a dynamics. Our theoretical description\cite{DKRT2003,DKRT2004,KMRST2001} is based on the use of a non-ergodic, deterministic, {\it zero-temperature}-like dynamics which neglects all thermally activated processes that would allow the system to cross the (grand-potential) barriers: a change in the fluid configurations is thus only due to a change in the landscape that comes itself from a change in the temperature or from a variation of the external field (here the chemical potential $\mu$).  As $\mu$ varies (for instance, when the pressure in the reservoir is slowly increased at constant $T$), the system  either follows the minimum in which it was trapped as this minimum deforms gradually, or it falls instantaneoulsy into another minimum when the former reaches its stability limit. This latter move  is a discontinuous
and irreversible process, an  {\it avalanche}, which is at the origin of the history-dependent behavior of the system. The avalanche corresponds to some {\it collective} condensation event inside the gel network which manifests itself by a jump in the adsorption isotherm (whereas the  gradual deformation of a minimum only corresponds to a slight swelling of  the liquid domains). The size of the avalanche is thus given by the size of the jump, i.e. the discontinuous change in the average fluid density $\rho_f=(1/N) \sum_i \rho_i$. As was shown qualitatively in Ref.\cite{DKRT2004}, these condensation events occur at different length scales inside the gel, depending on chemical potential, temperature, and porosity.  In particular, when $\phi$ is large enough, for instance  $\phi=95\%$, there exists a critical point separating a high-temperature regime where all condensation events are of microscopic or mesoscopic size (and the adsorption isotherm is smooth in the thermodynamic limit), and a low-temperature regime where a  macroscopic fraction of the gas condenses, which yields a finite jump in the isotherm.
It is one goal of the present work to put these observations on a more quantitative basis. 

We shall discuss below in more detail  the relevance of this theoretical description to the actual experimental situation. Let us just recall that this picture is  expected to be valid when there are widely separated time scales in the problem. Specifically, the  observation time must be much larger than the local equilibration  time  (i.e. the time needed to relax from a marginally stable state to a nearby metastable state) but  much smaller than the time associated to thermal activation (i.e. the time that it takes to escape from a local minimum via a nearby saddle point)\cite{note5}.  This requires that energy barriers are much larger than temperature, which is indeed the case in many experimental situations dealing with random systems. It may also be noted  that our definition of avalanches generalizes the one used at $T=0$ (see e.g. Ref.\cite{SDM2001}).  We indeed  introduce an explicit temperature-dependence in the problem by allowing  the (mean-field) free-energy landscape itself to change with $T$ (for a similar approach in the context of random magnets, see e.g. Ref. \cite{SLG1983}). As shown in our previous works, this dependence explains quite convincingly the variations observed in the hysteretic behavior of gases adsorbed in disordered solids at low enough temperature. 

 \section{Avalanche properties}
 
As discussed in the preceding papers of this series\cite{DKRT2003,DKRT2004}, the number, size and shape of the condensation events along an adsorption isotherm depend on porosity and temperature. They  also depend on the value of the chemical potential at which they occur. The random character of these events (each finite-size sample has a  different isotherm) requires a statistical study of their properties, an issue that has been investigated  in great detail in the context of the field-driven RFIM\cite{PDS1996, PV2003}. 

\subsection{Numerical procedure}

The first problem is to identify each avalanche properly. At $T=0$, when the microscopic variables can only take discrete values (such as the Ising spins in the RFIM), this is an easy task because the energy landscape only changes through the disappearance of minima, which induces the avalanches. One can then vary the external field until a spin becomes unstable (this spin is unique when the random field distribution is continuous) and then count the number of spins that flip during the avalanche to determine its size.
At finite temperature, on the other hand, one has to deal with continuous variables (the local fluid densities $\rho_i$ in the present case), and since  $\delta \mu$, the increment in the chemical potential along the adsorption isotherm, is always finite, one has to distinguish between the gradual deformation of a minimum and the  jump associated to the passage to  another local minimum, both leading to a change in the total density. Moreover, if $\delta \mu$  is not small enough, several avalanches may overlap, which biases the statistics of their number or size distribution.
Although, in principle, one could precisely locate the beginning of each avalanche by following the evolution of the eigenvalues of the Hessian matrix $H_{ij}=\partial ^2\Omega/ \partial \rho_i \partial \rho_j$ with $\mu$ or $T$ (an avalanche starts when a local spinodal corresponding to the vanishing of a certain eigenvalue is reached\cite{note6}), this is a daunting numerical task for a large system (with $N\sim 10^6$) and it is more advisable to use another procedure.  

Let us first note that the determination of the location and size of the avalanches depends on the precision in the computation of the metastable states. Since these latter are obtained by solving the set of Eqs. (2) iteratively,  it must be checked that the results are not significantly modified when the convergence criterium of the iterative procedure is changed. All the results presented here were computed with a precision $\epsilon=10^{-6}$ (i.e., the calculation was stopped at  iteration $n$ when $|\rho_i^{(n)}-\rho_i^{(n-1)}|<\epsilon$). It was found that decreasing $\epsilon$ to $10^{-10}$ did not change the results by more than $1\%$ (but in turn the computational time was increased by a considerable factor).

\begin{figure}[hbt]
\includegraphics[width=8cm]{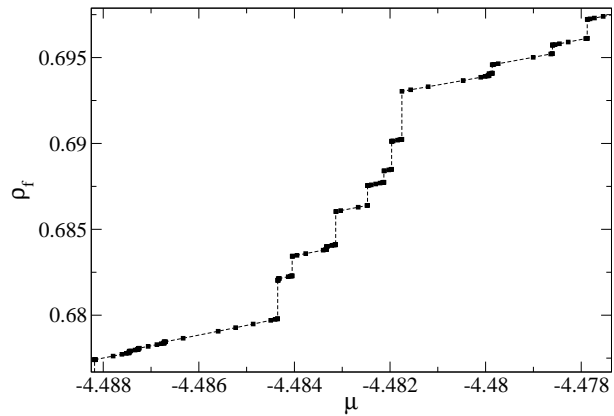}
  \caption{Portion of a typical adsorption isotherm in a $87\%$ porosity aerogel at $T^*=T/T_c=0.5$ ($\rho_f$ is the average fluid density inside the aerogel  and $\mu$ the gas chemical potential). Calculations were done in a sample of size $L=50$ using the dichotomy procedure explained in the text.}
\end{figure}

Consider now the variations $\delta\rho_i=\rho_i(\mu+\delta \mu) -\rho_i(\mu)$ in the local fluid densities when increasing the chemical potential by a very small (but finite) increment $\delta \mu$ (for instance $\delta \mu= 10^{-6}$, with $w_{ff}$ taken as the energy unit). We show in Fig. 1 a small portion of a typical low-temperature isotherm computed in a $87\%$ porosity sample of size $L=50$ at $T^*=T/T_c=0.5$ ($k_BT_c/w_{ff}=2$). The random character of the condensation events results in the fine staircase structure of the curve, with the vertical steps corresponding to the  avalanches separated by nearly horizontal portions corresponding to the smooth deformation  of the grand-potential local minima (note the small scale on the vertical axis). The slope of these near-plateaus depends on temperature and its order of magnitude can be estimated from the isothermal compressibility of the pure lattice gas. It turns out that {\it all } the $\rho_i$'s vary by a very small amount in these flat portions of the isotherms (less than $10^{-5}-10^{-6}$) whereas the vertical steps are due to a significant variation of {\it some} of the $\rho_i$'s (a low temperature, this typically corresponds to a gazeous region becoming liquid). We can use this feature  to identify the avalanches unambiguously. Specifically, we consider that the change $\delta \rho_f=\sum_i \delta\rho_i$ in the isotherm is only due to a smooth deformation of a  grand-potential minimum when $\delta\rho_{\max}=\max_i \{\delta \rho_i\}$ is smaller that some threshold, which we take equal to $0.1$. Otherwise there is an avalanche, and we associate to it all the sites for which $\delta\rho_i>0.001$.  The size of the  avalanche is then $s=\sum_{i,av}\delta\rho_i$, where the sum is restricted to these sites\cite{note20} (by taking $s=\delta \rho_f=\sum_i \delta\rho_i$  one would neglect the small increase of $\rho_f$ which is due to the ``smooth" variation of the $\rho_i$'s on the other sites of the lattice).  This criterium  thus implies that avalanches for which $\delta\rho_{\max}<0.1$ cannot be detected. By choosing a smaller threshold, e.g. $\delta\rho_{\max}<0.01$, one may find new avalanches, but their size $s$ is very small (always less than $10$). As a consequence, avalanches with $s<10$ will not  be considered in the following. In any experiment there is also a threshold below which signals cannot be detected.

Since one must collect the results of many samples to get a good statistics of the avalanches properties, it is impossible in practice to study a whole isotherm with an increment $\delta \mu$ as small as $10^{-6}$.  It is much more efficient to proceed by dichotomy, comparing the configurations obtained for  two values of $\mu$, say $\mu_1$ and $\mu_2$ (initially, $\mu_1$ is a large negative value and $\mu_2=-4$, the value at saturation). If  $\delta\rho_{\max}>0.1$ between the two configurations, one then compares the configurations for  $\mu_1$ and  $(\mu_1+\mu_2)/2$ on the one hand, and $(\mu_1+\mu_2)/2$ and $\mu_2$ on the other hand. One then looks for an avalanche in the two intervals, and the same procedure is repeated until no avalanche is found between two configurations, or the difference in chemical potential becomes smaller than $10^{-6}$ (this is the method that was used for obtaining the isotherm shown in Fig. 1, which explains that the points are irregularly spaced).

Finally, to characterize geometrically an avalanche, which is a random object, one may define an effective ``radius of gyration" $R_g$  by attributing  to each site $i$ a ``mass" that corresponds to the change in the local density,
\begin{align}
R_g^2=\frac{\sum_{i,av}({\bf r}_i-{\bf r}_m)^2\delta \rho_i }{\sum_{i,av}\delta \rho_i }
\end{align}
where the sum is restricted to the sites contributing to the avalanche (i.e. with $\delta\rho_i>10^{-3}$) and ${\bf r}_m=\sum_{i,av}{\bf r}_i \delta \rho_i /\sum_{i,av}\delta \rho_i $ is the ``center of mass" of the avalanche\cite{note7}.

\subsection{Number, size and shape}

Let us first study how the number of avalanches with a given size varies with temperature and porosity. To illustrate the general trend, we consider  isotherms obtained in $87\%$ and $95\%$  porosity aerogels, as shown in  Figs. 2(a) and 3(a).  The sequence of avalanches was recorded in  $500$  samples of linear size $L=50$ and  $50$ samples of linear size $L=100$, respectively (this approximately corresponds to  the same ratio $L/ \xi_G\approx 10$, where $\xi_G$ is the correlation length of the aerogel\cite{DKRT2003}). 
Figs. 2(b) and 3(b)  show $N(s)$, the (unnormalized) avalanche size distribution ($\int N(s) ds=N_{av}$, the total number of avalanches of size larger than $10$). As $T$ increases, $N_{av}$ decreases from $335000$ to $81000$ and from $83700$ to $19800$, respectively. The size of the largest possible avalanche corresponding to a simultaneous change of all the $\rho_i$'s from $0$ to $1$ is  $2L^3$. One thus has $s_{max}=2.5 \ 10^5$ and $2. \ 10^6$ in the $87\%$ and $95\%$  aerogels, respectively.

It is readily seen  that there exist much larger avalanches in the lighter aerogel. This is a direct  consequence of the larger void space available, as discussed in our previous papers\cite{DKRT2003,DKRT2004}. In particular, at $T^*=0.5$, some avalanches span the whole system, signaling that a macroscopic avalanche possibly occurs in the thermodynamic limit. Note however that the statistics for very large avalanches is not good, which explains the erratic behavior seen in Fig. 3(b) for large $s$. Moreover, finite-size effects play an important role in the large-porosity/low-temperature regime and they should be taken into account to properly describe  the evolution towards the thermodynamic limit (see Ref.\cite{PV2003} for a careful study in the case of the RFIM). Unfortunately, this would require a considerable amount of numerical work, and we shall therefore only concentrate our discussion on the non-spanning avalanches. 
\begin{figure}[hbt]
\includegraphics*[width=10cm]{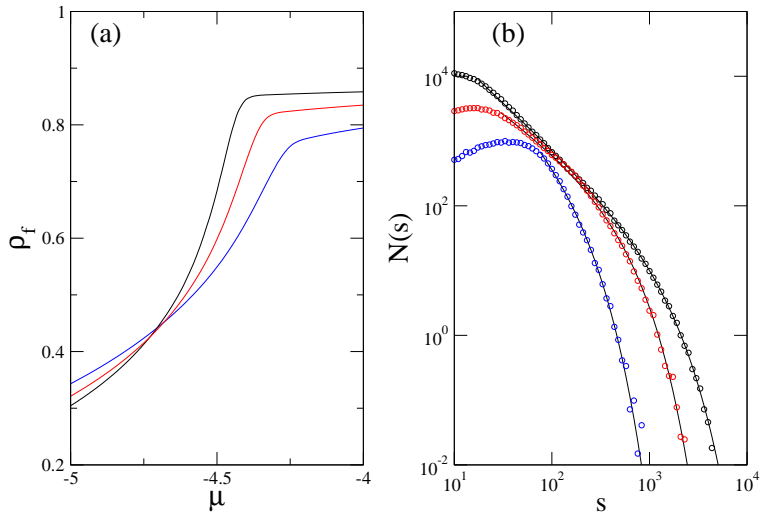}
  \caption{(a) Average isotherms in a $87\%$ porosity aerogel at $T^*=0.5, 0.65 $,  and $0.8$ (from top to bottom); data correspond to an average over $500$ gel samples of size $L=50$. (b)Corresponding (unnormalized) avalanche size distributions $N(s)$. The solid lines represent the fit $N(s) \sim s^{-\tau} e^{-s/s_0}$ (Color on line).}
\end{figure}

\vspace*{0.6cm}

\begin{figure}[hbt]
\includegraphics*[width=10cm]{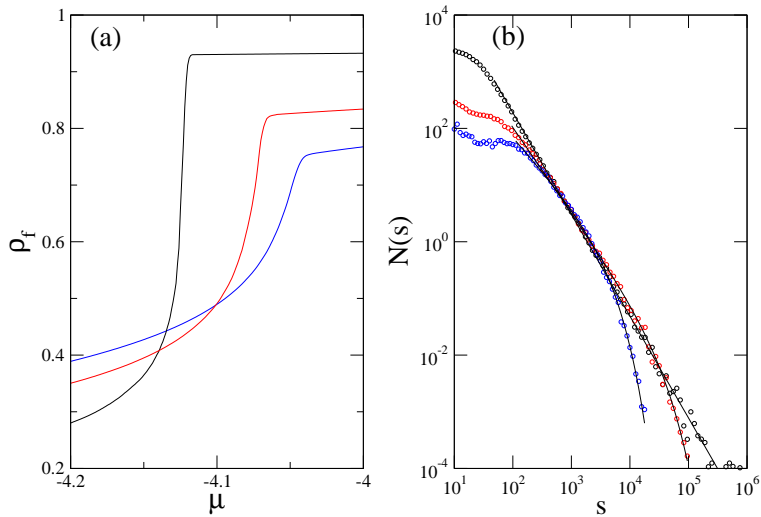}
  \caption{Same as Fig.2 in a $95\%$ porosity aerogel at $T^*=0.5, 0.8$, and $0.9$; data correspond to an average over $50$ gel samples of size $L=100$ (Color on line).}
\end{figure}
Figs. 2(b) and 3(b) show that the evolution of $N(s)$ with temperature is quite similar in the two aerogels: as $T$ increases, both large {\it and} small avalanches are suppressed, and, as a consequence, the total number  $N_{av}$ decreases. Only the number of avalanches of intermediate size  stays roughly constant with $T$.
If one refers to what happens in the field-driven RFIM at $T=0$, this result may appear surprising. Indeed, the isotherms get smoother as one increases the temperature,  like the magnetization curves $M(H)$ in the RFIM as one increases the width  of the random field distribution\cite{S1993,PDS1996}. This seems to imply that the temperature plays the role of an effective disorder in the system. Therefore, one would expect that large avalanches become rare as $T$ increases (which is indeed what happens), but  also that small avalanches proliferate, so that $N_{av}$ increases on  average\cite{PDS1996,PV2003}. It turns out, however, that temperature also plays another role in our system, a role which is well documented  in the literature devoted to capillary condensation\cite{E1990}: adsorption in a small cavity becomes more gradual as $T$ increases. Specifically, mean-field calculations for a single pore of simple geometry (for instance a slit of width $l$) predict that adsorption is continuous and reversible  above some critical temperature $T_c(l)$ which decreases as $l$ decreases.  Since small avalanches are associated with gas condensation in small cavities or crevices defined by the neighboring gel strands, this phenomenon explains the decrease of the number of small avalanches (when $T>T_h(\phi)$, the temperature at which the hysteresis disappears, there are no avalanches at all and the isotherms become reversible\cite{DKRT2004}).  It is clear that  the  random-field picture can only be applied at a scale much larger than the correlation length of the solid: schematically, capillary condensation prevails for  $s^{1/3}<\xi_G$ whereas the disordered character of the aerogel structure plays a major role  for $s^{1/3}>\xi_G$. The two mechanisms compete at the scale $s ^{1/3}\sim \xi_G$ which explains that the number of avalanches of intermediate size does not vary significantly with $T$. 

If we now concentrate on the disorder-dominated  regime, the most remarkable feature, quite  visible  in the $95\% $ gel at $T^*=0.5$, is that $N(s)$ follows an approximate power-law behavior over several decades.  This is  the defining signature of criticality, an issue that has been at the focus of intense research activity in recent years, for instance to explain the scaling properties of the Barkausen noise in magnetic materials\cite{DZ2004}.  In the RFIM picture, the power-law behavior  is  related to the existence of a disorder-induced critical point\cite{S1993,PDS1996}. In this context, the avalanche size distribution should be analyzed using the  form 
\begin{align}
N(s) \sim s^{-\tau} \Phi(s/s_0)
\end{align}
where $\tau$ is a critical exponent\cite{note15}, $\Phi$ is a scaling function, and $s_{0}$ a cut-off that depends on the distance to the critical point ($s_0$ diverges as the critical disorder is approached). Our results are however too limited to obtain sensible scaling collapses (although a crude attempt shows  that the scaling function is non-monotonic, in agreement with the RFIM behavior\cite{PDS1996}). We simply replace the function $\Phi$ by an exponential form only describing the cut-off of the power-law at the scale $s_0$.  Figs. 2(b) and 3(b) show the fits of $N(s)$ in the two aerogels according to this approximate treatment. The power-law regime  is very limited in the $87\%$ aerogel even  at the lowest temperature, indicating that  the system is clearly outside the critical region. On the other  hand, in the $95\%$ gel, the curves  show a  linear portion on a log-log scale that increases as $T$ decreases: the cutoff $s_0$ is pushed to larger values of $s$ (from $3200$ to $29500$ as $T^*$ varies from $0.9$ to $0.8$) and reaches the box size for $T^*=0.5$ (our set of data is however too limited to estimate the critical exponent associated to the divergence of $s_0$).  The exponent $\tau$  increases and becomes approximately equal  to $1.8$ in the fully developed power-law regime at  $T^*=0.5$.  Taking into account the fact that our systems are small  and that $\tau$ in a finite system depends on disorder\cite{PV2003}, this value does not appear inconsistent with that predicted for the RFIM, $\tau=2.03\pm0.03$\cite{PDS1996}. In any case, the gel-fluid system is clearly close to criticality at this temperature.

\begin{figure}
\includegraphics[height=7cm]{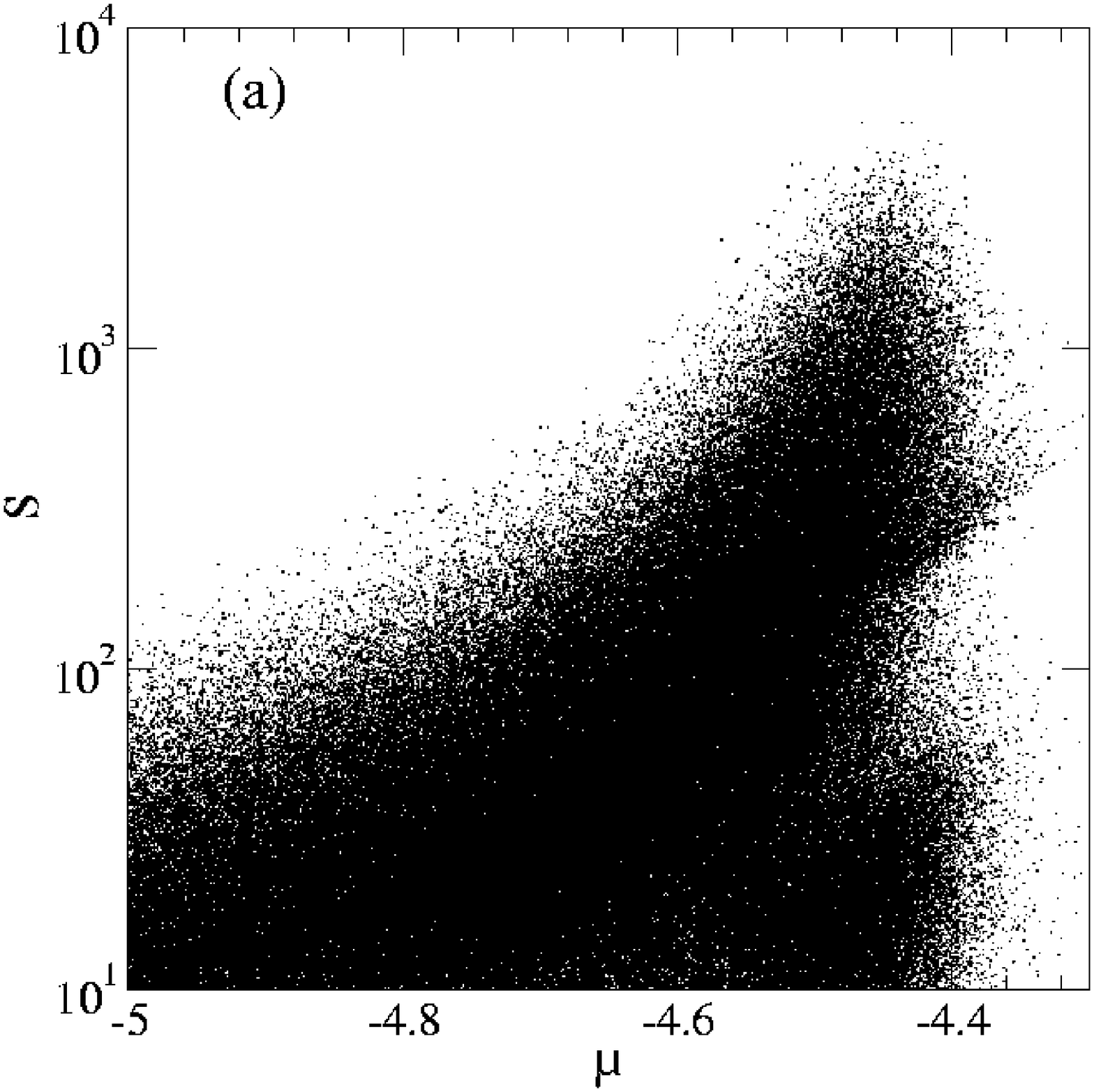}
\includegraphics[height=7cm]{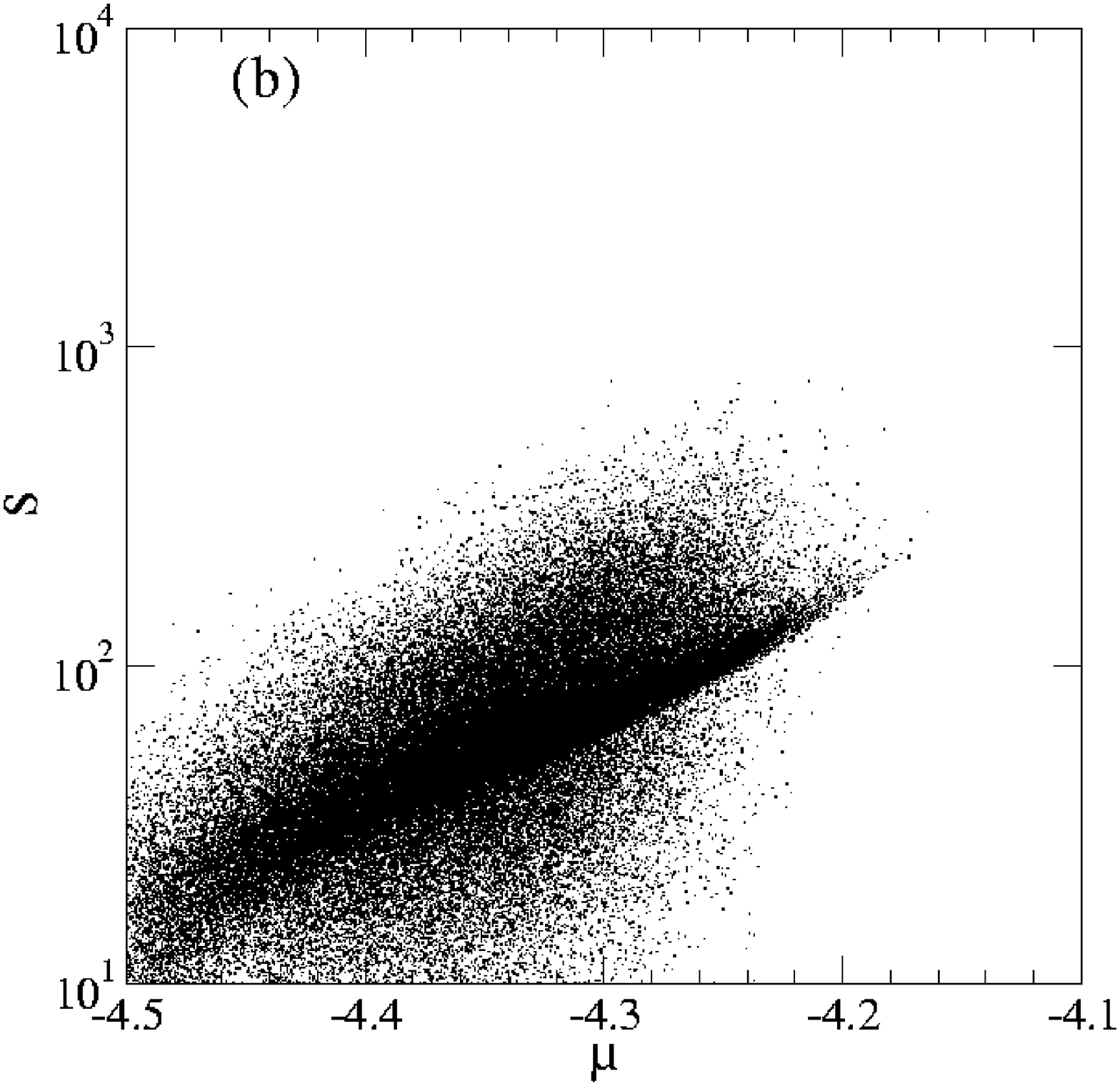}
 \caption{Point cloud of the avalanche size $s$  recorded in the  $87\%$ porosity samples as a function of chemical potential $\mu$ at $T^*=0.5$ (a)  and $T^*=0.8$ (b)}
\end{figure}

\begin{figure}
\includegraphics[height=7cm]{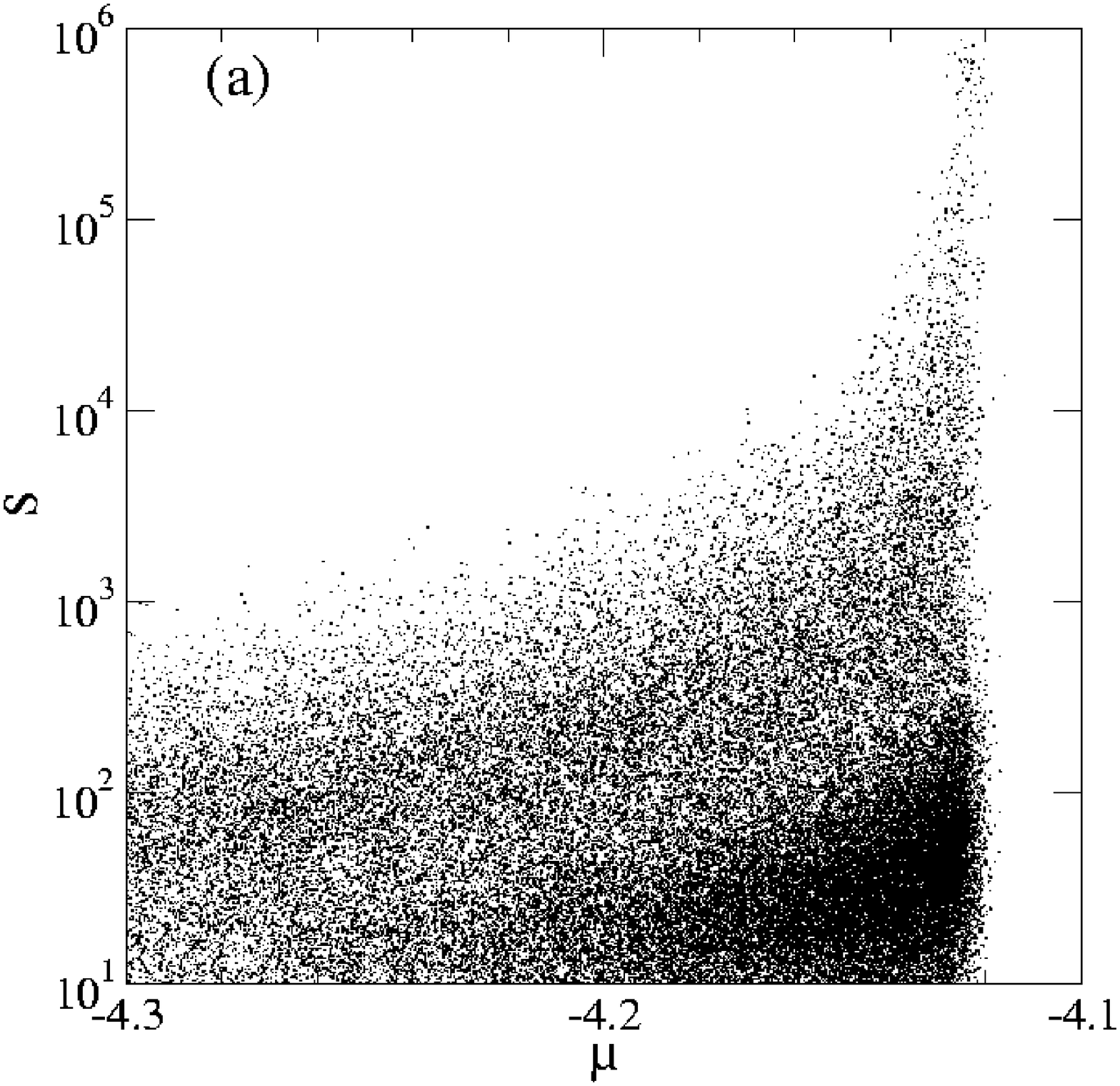}
\includegraphics[height=7cm]{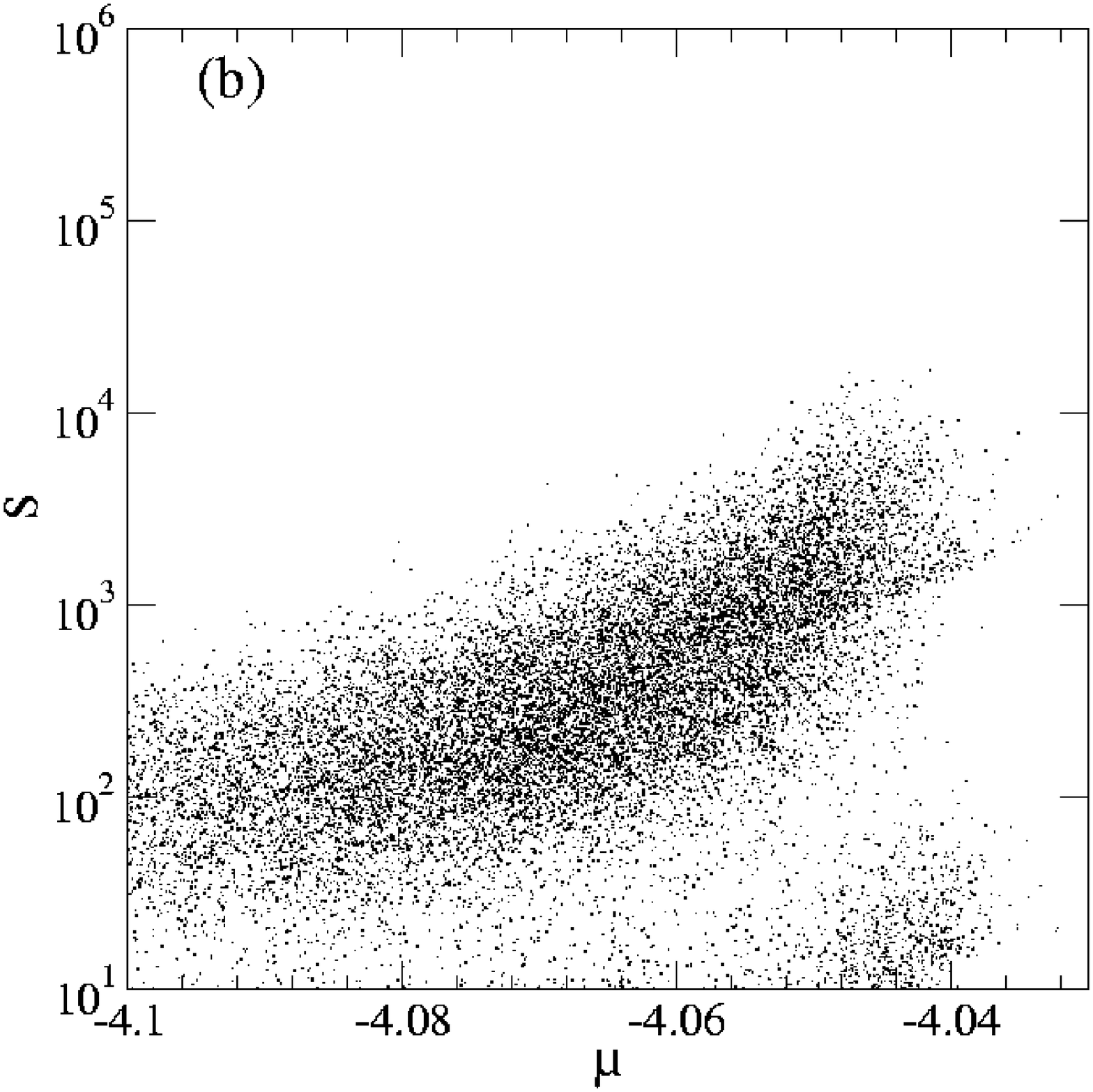}
\caption{Same as Fig. 4 in the  $95\%$ porosity samples at $T^*=0.5$ (a) and $T^*=0.9$ (b).}
\end{figure}

The number and the size of the avalanches along an adsorption isotherm strongly depend on the value of the chemical potential.  We thus also recorded the value of $\mu$ at which each avalanche of size $s$ occured.  Figs. 4 and 5  show  the corresponding point clouds in the $87\%$ and  $95\%$ gels at $T^*=0.5$ and $0.8$, and $T^*=0.5$ and $0.9$, respectively. As a general rule, the size of the largest avalanches increases with $\mu$, especially in the last stage of the adsorption (note the logarithmic scale on the vertical axis). But the most striking feature is that avalanches of very different sizes occur in a small interval of $\mu$.  For instance, in the $87\%$ gel at $T^*=0.5$ and $\mu \approx -4.5$, $s$ varies from $10$ to more than $10^3$. In the $95\%$ gel at the same temperature and $\mu \approx -4.12$, the avalanches span almost all possible sizes between  $10$ and $10^6$. These observations are in complete contradiction with the picture based on the independent pore model which is often used to interpret  capillary condensation phenomena\cite{GS1982}. In this model, only pores of a certain characterisic size $l$ fill at a given value of the chemical potential, and the adsorbed fluid density is directly related to the number of pores of this size that can be found in the material. (One then try to extract from the isotherm the pore size distribution.) This picture is clearly wrong in the present case because of the very open and intricate character of the gel structure that allows condensation events of very different sizes.

The geometrical characterization of the avalanches is rather difficult. Using $s^{1/3}$ as an estimate of the  extension of an avalanche is misleading because events with very different shapes are observed (see for instance Figs. 6 and 7 in Ref.\cite{DKRT2004}). 
This variety in the geometry is reflected in the values of the radius of gyration $R_g$, as defined by Eq. (3).  The value of $R_g$ is indeed proportional to the maximal extension of the avalanche, but the prefactor depends on its shape.  Figs. 6 and 7 show how $R_g$ varies with $s$ in the $87\%$ and  $95\%$ aerogels. One can see that avalanches with a similar size may have quite different radii of gyration. For instance, in the $87\%$ aerogel at $T^*=0.5$, avalanches of size $s\sim 10^3$   have a radius that varies between $4$ and $10$ (note that for a compact object, a factor of $2$ in $R_g$  would correspond to a factor $8$ in $s$). Similarly,  in the $95\%$ aerogel at the same temperature,  $R_g$ varies from $10$ to $15$ for avalanches of size $s\sim10^4$. It is therefore impossible to associate a unique characteristic length to an avalanche of a given size.  Moreover, by examining the avalanches that occur in a small interval of $\mu$, we  find that they also have very different radii of gyration. 

As the temperature increases, the values of $R_g$ for a given size $s$ are less scattered. Indeed, avalanches become more compact as they correspond to condensation events that occur in a single cavity of the aerogel. This evolution is very clear in the $87\%$ aerogel: at $T^*=0.8$, the radius of gyration of most avalanches is between  $2$ and $4$ and there are almost no avalanches with $R_g$ larger than 6. On the other hand, there is still a large domain of variation of $R_g$ in the $95\%$ aerogel at $T^*=0.9$,  and one should go to even higher temperatures to observe a significant reduction. There is in fact  a remarkable resemblance between Figs. 6(a) and 7(b) which suggests that the avalanche properties are similar apart from a scale factor of about $2$.  The adsorption isotherms have also similar shapes, as can be seen in Figs. 2(a) and 3(a).  This confirms the observation made in Ref.\cite{DKRT2004} that the adsorption processes in a low porosity gel at low temperature and in high porosity gel at a higher temperature have similar properties. 

\begin{figure}
\includegraphics[height=7cm]{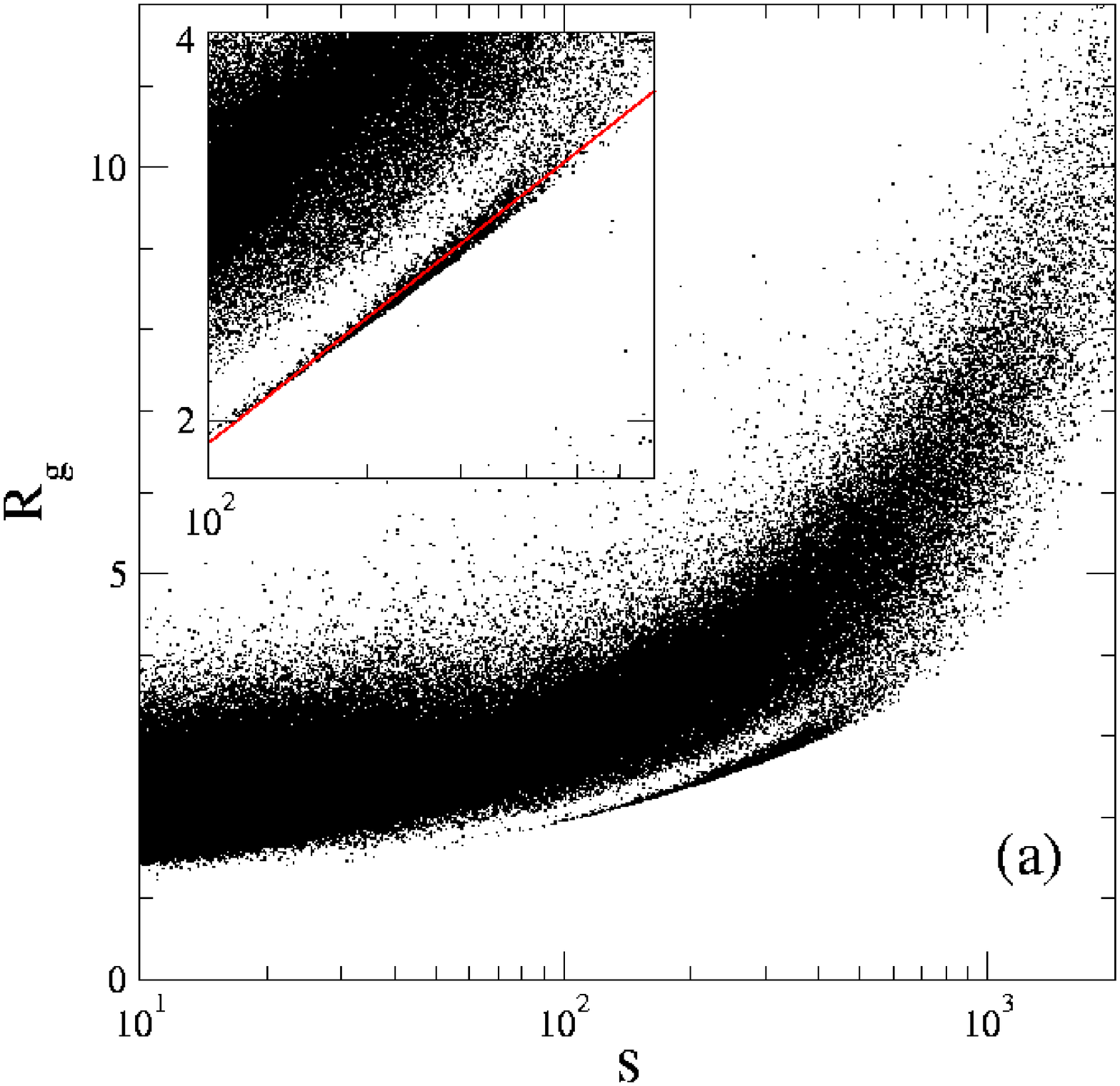}
\includegraphics[height=7cm]{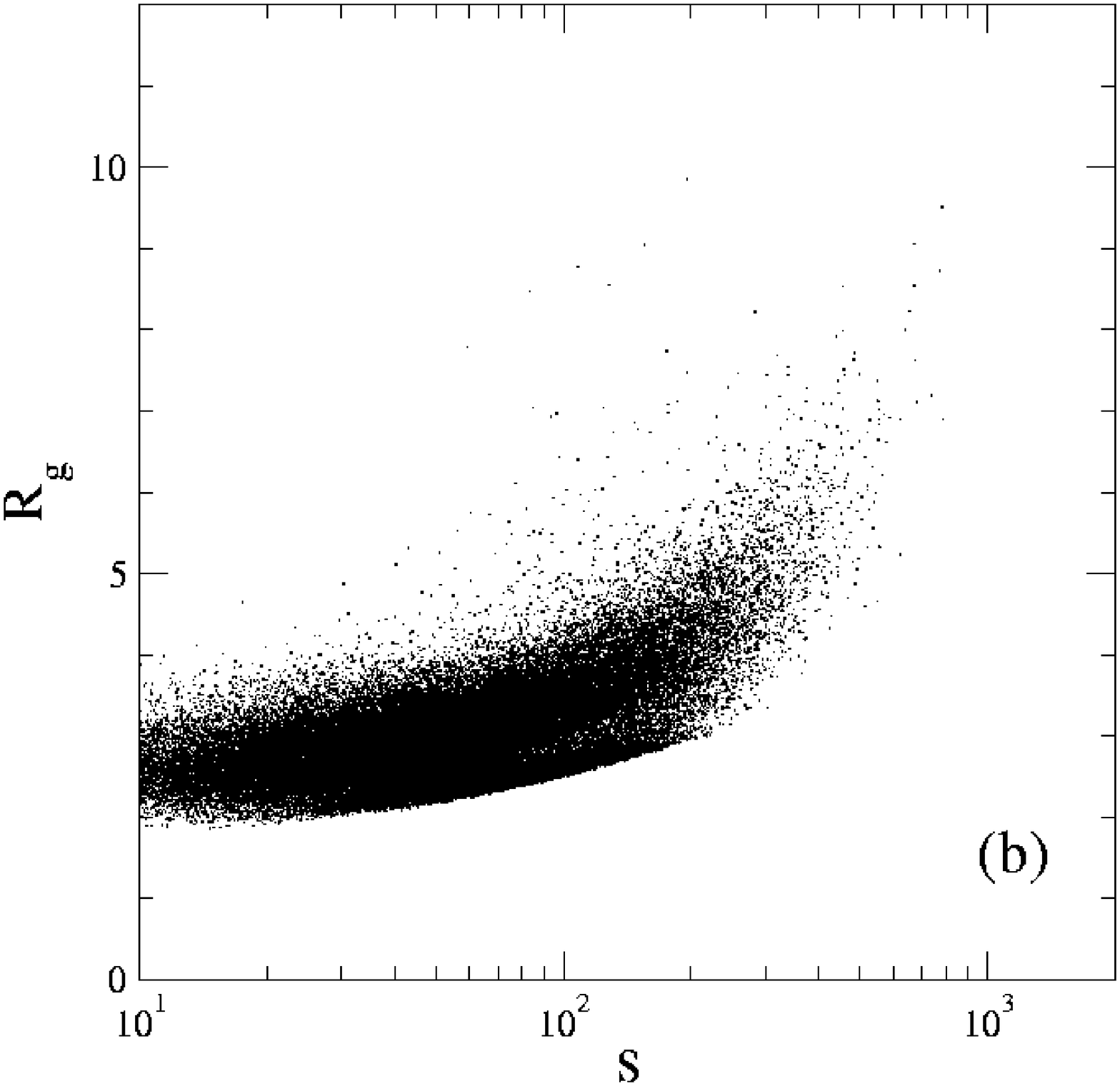}
  \caption{ Radius of gyration $R_g$ and size $s$ of the avalanches in the $87\%$  porosity samples  at $T^*=0.5$ (a) and $T^*=0.8$ (b). The inset in (a) is a magnification  of the point cloud of the spherical avalanches (see text)  in log-log scale; the solid line represents the equation  $R_g=0.42\  s^{0.33}$ (Color on line).} 
\end{figure}

\begin{figure}
\includegraphics[height=7cm]{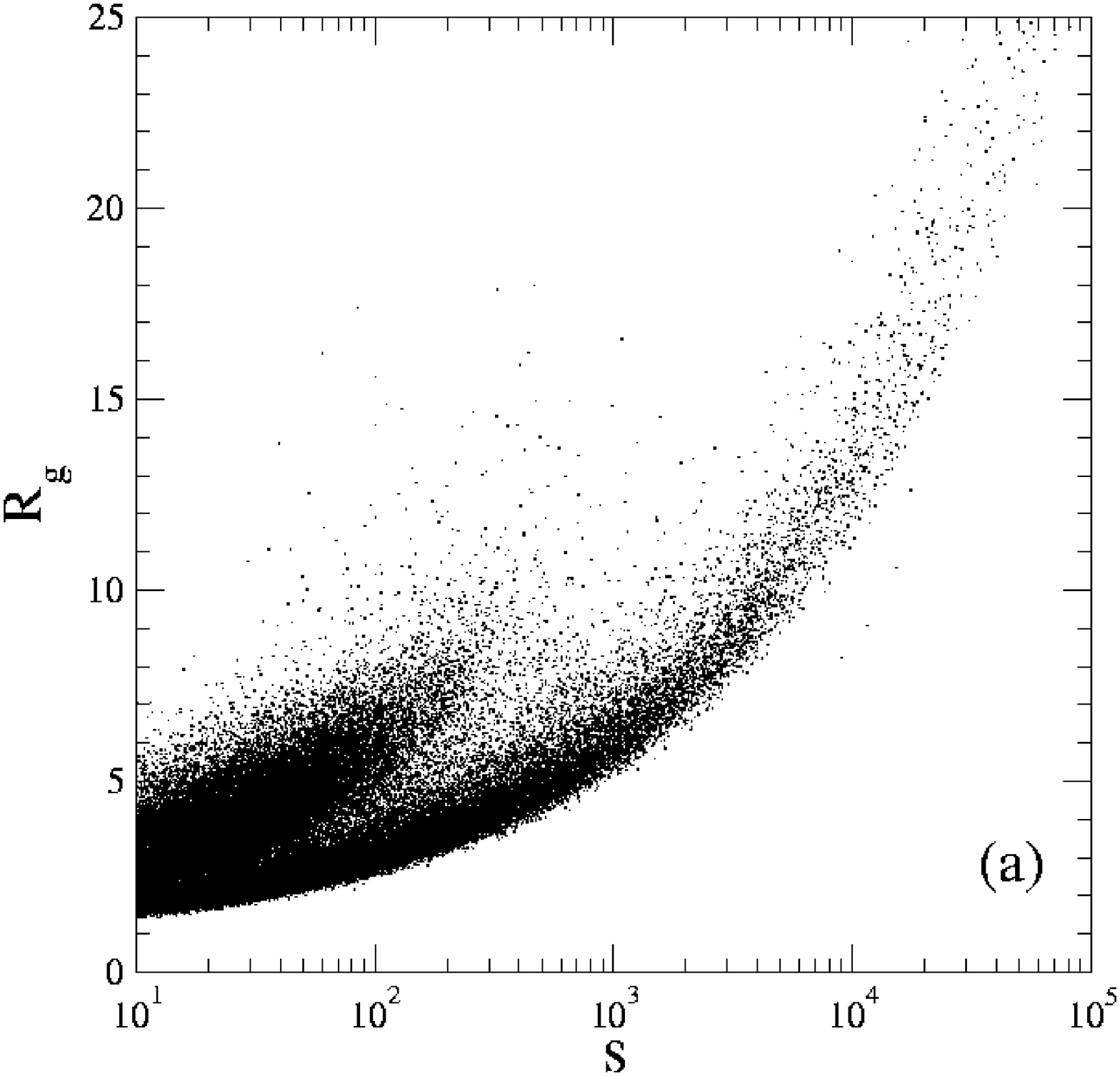}
\includegraphics[height=7cm]{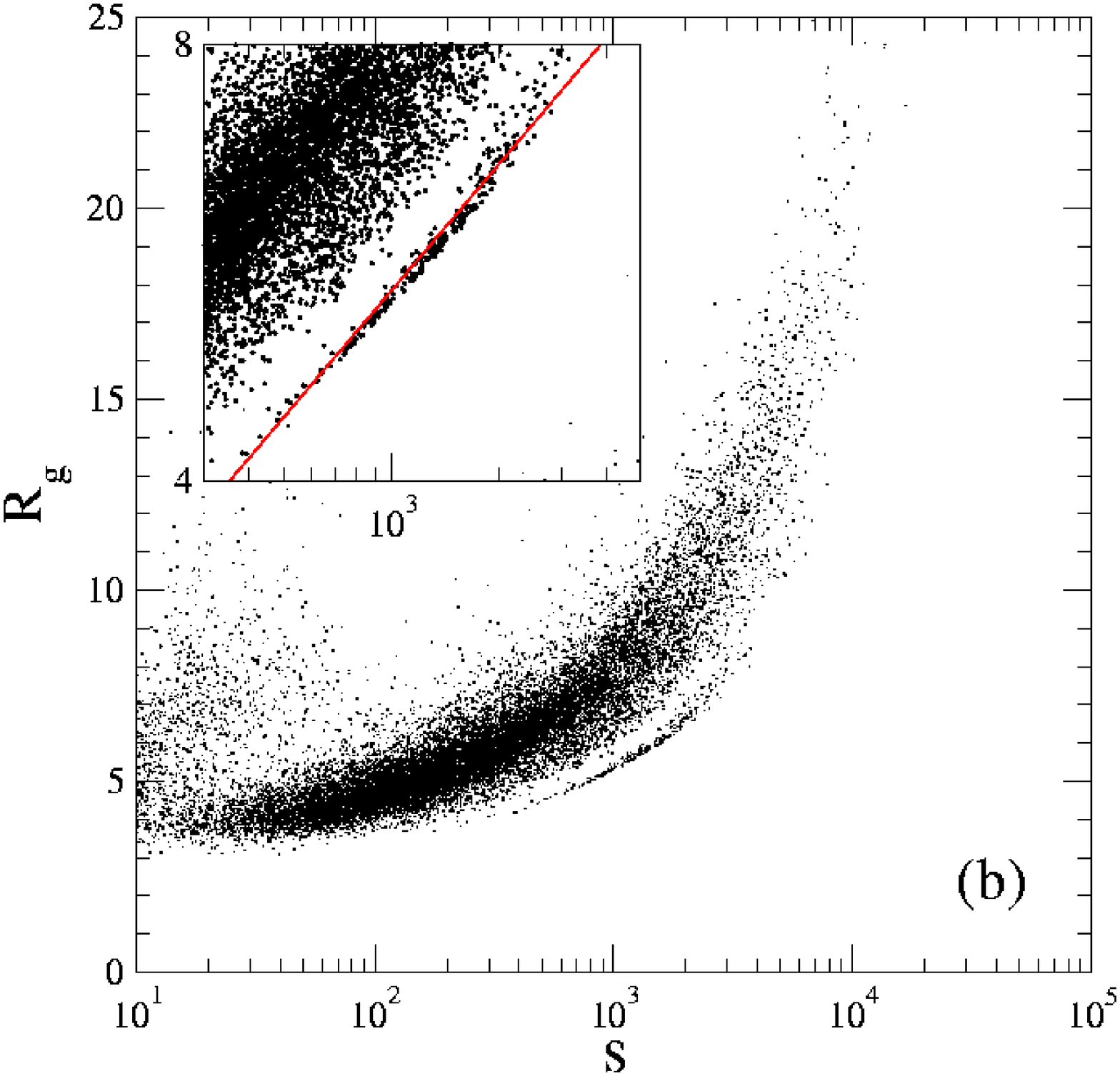}
  \caption{Same as Fig. 6 for the $95\%$ porosity samples at $T^*=0.5$  and $0.9$. In the inset of (b) the solid line represents the equation  $R_g=0.73\  s^{0.29}$ (Color on line).}
\end{figure}

\subsection{Spherical avalanches}

In both Fig. 6(a) and Fig. 7(b), one may notice that a small set of avalanches stands out . In the $87\%$  gel, at $T^*=0.5$,  these avalanches have a size $s$ in the range $100-500$ and a smaller radius of gyration than the other ones. They are therefore more compact, and a direct observation of some of them shows that their shape is almost perfectly spherical. This  is confirmed by a systematic study of the asphericity parameter which is  always smaller than $10^{-2}$\cite{note12}.  As shown in the inset of Fig. 6(a) on a  log-log scale, the radius of gyration of these avalanches varies approximately like $R_g \approx 0.42 \ s^{0.33}$ which is indeed close to the expected value for spherical compact objects ($R_g=\sqrt{3/5}\ R=\sqrt{3/5} \ [3s/(4\pi)]^{1/3}\approx 0.48 s^{1/3}$ for a sphere of radius $R$). The radius $R$ varies between $2.5$ and $4$, this latter value corresponding approximately to $\xi_G$ which is also more or less the size of the largest voids in the aerogel \cite{DKRT2003}. Moreover, as shown in Fig. 8, most of these avalanches occur at the end of the adsorption process: they thus correspond to the condensation of the last gas bubbles that remain in the large cavities. 
\begin{figure}
\includegraphics[width=6cm]{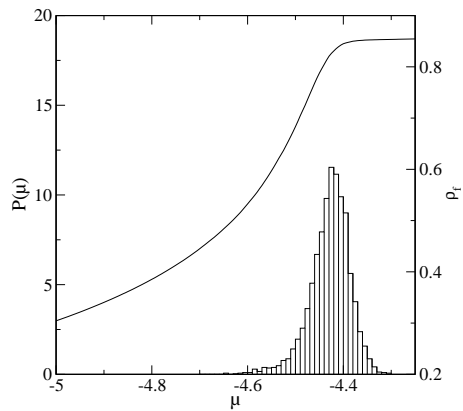}
  \caption{Probability distribution $P(\mu)$ of spherical avalanches along the adsorption isotherm in the $87\%$ porosity aerogels at $T^*=0.5$. The average isotherm is shown as a continuous line.}
\end{figure}

A similar behavior is observed in the $95\%$ aerogel at $T^*=0.9$ (Fig. 7(b)).  The avalanches are again approximately spherical and occur in the last stage of the adsorption process. However, since the temperature is higher, the boundaries  of the gas bubbles are less well defined, and the  radius of gyration of varies approximately like $R_g \approx 0.73 \ s^{0.29}$. The largest observed value of $R=\sqrt{5/3} \ R_g$ is  again comparable to the size of the largest cavities in the aerogel ($\xi_G\approx 10$).
Although there are also some other remaining bubbles of more complicated shape (see Figs. 7 and 9 in Ref.\cite{DKRT2004}), these spherical avalanches actually represent the only case where one  can relate directly the characteristic size of the condensation events to the structure of the aerogel. We stress again that, in general, the avalanches have a complicated shape  so that one cannot describe the condensation of the adsorbed gas  in terms of ``constant curvature bubbles"\cite{TYC1999}.

\vspace*{0.2cm}
\begin{figure}[hbt]
\includegraphics*[width=9.5cm]{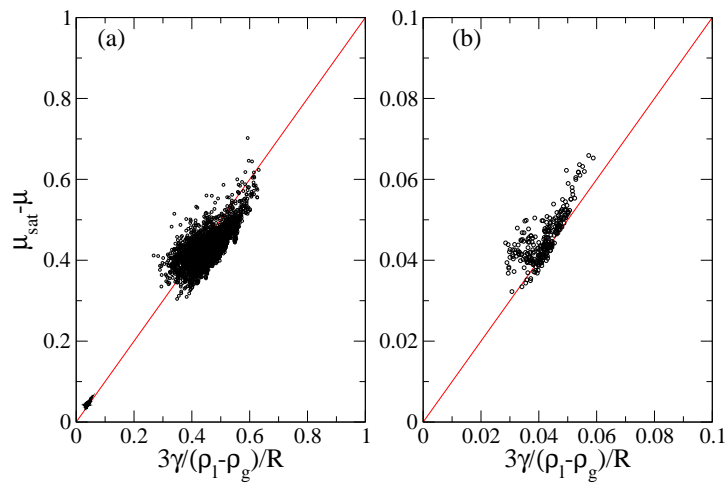}
  \caption{(a) The undersaturation $\Delta \mu=\mu_{sat}-\mu$  as a function of the inverse radius  of the spherical avalanches in the  $87\%$ and $95\%$ aerogels at $T^*=0.5$ and $T^*=0.9$, respectively. (b) is a magnification of the small point cloud in (a) near the origin  corresponding to the avalanches in the  $95\%$ aerogel. The solid line with a slope $1$  corresponds to the Kelvin equation (see text) (Color on line).}
\end{figure}
The fact that spherical avalanches are clearly identified gives us the opportunity for testing a macroscopic description of the gas condensation in terms of a competition between volume and surface contributions. Such a description underlies the classical interpretation of capillary condensation based on the Kelvin equation\cite{GS1982} that is commonly used to extract from the isotherms information on the characteristic pore size in the solid. We thus write the change in the grand potential associated to the condensation of a spherical gas bubble of radius $R$  at the chemical potential $\mu$ as\cite{EMP1986}
\begin{equation}
\Delta \Omega(\mu)= [p(\mu) -p^+_l(\mu)] \frac{4\pi}{3}R^3-\gamma 4\pi R^2
\end{equation}
where $p(\mu)$ is the corresponding bulk pressure, $p^+_l(\mu)$ is the  pressure of the  metastable liquid at the same  chemical potential (both pressures are computed from the equation of state of the bulk lattice gas),  and $\gamma$ is the (planar) liquid-gas surface tension. All these terms except $R$ are temperature dependent. Assuming that condensation  occurs at equilibrium, the Kelvin equation (in its Laplace form) corresponds to setting $\Delta \Omega=0$ in the equation. Expanding as usual $p$ and $p^+_l$ about $p_{sat}$ and to first-order in $\Delta \mu=\mu_{sat}-\mu$ (which is valid for small undersaturations), this yields $\Delta \mu=3\gamma/[(\rho_l-\rho_g)R]$, where $\rho_l$ and $\rho_g$ are the densities of the coexisting liquid and gas at this temperature.  Since  the surface tension on a lattice depends on the orientation of the surface with respect to the lattice vectors, we estimate $\gamma$ by taking the average of the three values calculated in the principle lattice planes $(100)$, $(110)$ and $(111)$\cite{SG2004}.   The corresponding ``Kelvin plot" is shown in Fig. 9.  We see that the ``macroscopic" Kelvin equation provides  the correct trend for  the variation of $\Delta \mu$ as a function of $1/R$, in spite of the fact that actual condensation does not occur at equilibrium ($\Delta \Omega\neq 0$\cite{note13}) and that the adsorbent potential and the curvature corrections are neglected in Eq. (4) (note that $\Delta \mu$ is ten times smaller in the lighter gel, which is due to the effect of temperature, correctly described by the term $3\gamma/(\rho_l-\rho_g)$, and to the change in the average radius $R$ by a factor of about $2$). The scatter in the data, however, shows that the Kelvin equation is only an approximation.

\section{Scaling of the adsorption isotherms and phase diagram}  

The existence of a critical point in our system, at which an infinite avalanche occurs for the first time in the thermodynamic limit and the avalanche size distribution has a power-law behavior, suggests the possibility of performing also a  scaling collapse of the adsorption isotherms.  In the case of the  nonequilibrium  RFIM at $T=0$ where the disorder is controlled by  the width $\sigma$ of the distribution of the random fields, standard renormalization group arguments suggest that the magnetization curve $M(H)$ for different values of  $\sigma$ may be collapsed using the scaling form\cite{PDS1996}

\begin{align}
M(\sigma,H)-M_c(\sigma_c,H_c )\sim  \mid r\mid^{\beta} {\cal M}_{\pm}(h/\mid r\mid^{\beta \delta})
\end{align}
where $M_c$ is the critical magnetization (i.e. the magnetization at the critical field $H_c$ for the critical disorder $\sigma_c$), $r=(\sigma_c-\sigma)/\sigma_c$ and $h=H-H_c$ are the reduced disorder and reduced field respectively and ${\cal M}_{\pm}$ is a universal scaling function ($\pm$ refers to the sign of $r$). The estimated values of the critical exponents in three dimensions are $\beta =0.035 \pm 0.028$ and $ \beta \delta \simeq1.81\pm0.32$\cite{PDS1996}. 

\vspace*{0.7cm} 
\begin{figure}[hbt]
\includegraphics[width=9cm]{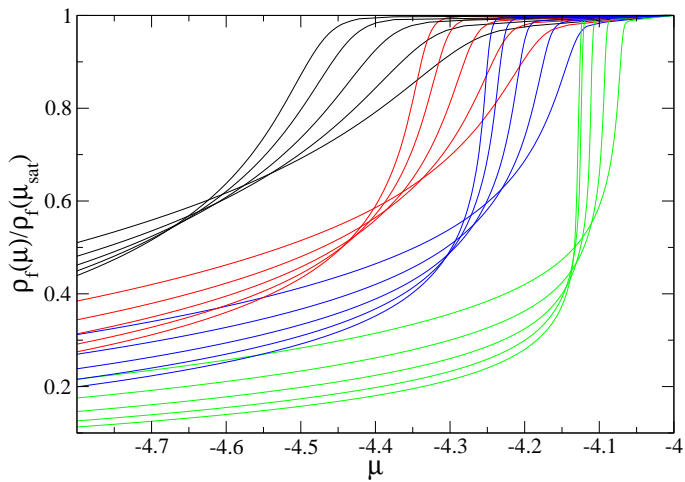}
  \caption{Average isotherms in aerogels of varying porosity (the fluid density $\rho_f$ is rescaled by the density at $\mu=\mu_{sat}$). From left to right: $\phi=87\%,90\%,92\%,95\%$ at $T^*=0.4, 0.5, 0.6,0.7,0.8$. The curves result from an average over $500$ samples of size $L=100$ (Color on line).}
\end{figure}
The situation in our system is more complicated than in the RFIM because the Hamiltonian (or, more correctly, the grand-potential functional $\Omega$) contains three parameters that can be tuned independently (the porosity $\phi$, the interaction ratio $y$, and the chemical potential $\mu$), and, moreover, the temperature controls the ruggedness of the free-energy landscape.  Although a rigorous proof is lacking, it seems however reasonable to assume that the critical behavior is governed by a unique fixed point, which may or may not be the same as for the $T=0$ nonequilibrium RFIM. In other words, there are only two  relevant variables from a  renormalization-group standpoint: the chemical potential (that plays the role of the magnetic field $H$ in the RFIM) and an effective disorder $\sigma$ which is some smooth funtion of $y$, $\phi$ and $T$. (In addition, one could also have to take into account the microstructure of the aerogel, on top of the dependence on the porosity: in particular, one may suspect  that the fractal character of very light aerogels, e.g. $\phi \geq 99\%$, modifies the values of the critical exponents or, at least, induces a different scaling behavior in some intermediate regime; this possibility is not taken into account in the present study  which does not consider aerogels of porosity larger then $95\%$.) For fixed $y$  and $T$, there is  a critical value  $\phi_c(y,T)$ of the porosity, and a critical value $\mu_c(y,T)$ of the chemical potential.
Alternatively, for fixed $y$ and $\phi$, there is a temperature $T_c(y,\phi)$ and a chemical potential $\mu_c(y,\phi)$ at which the system becomes critical. If this picture is correct, one expects to observe the same critical behavior whether the critical line  $\phi_c(T)$ (or $T_c(\phi)$) is approached along a constant-$\phi $ or a constant-$T$ trajectory  in the temperature-porosity diagram (for simplicity, we forget the dependence on $y$ which is kept constant and equal to $2$ in all our computations\cite{DKRT2003,DKRT2004}).  

Consider for instance the collection of isotherms displayed in Fig. 10 which covers a significant range of temperatures and porosities.  
This would correspond to $2.08K<T<4.16K$ for the real $^4$He-aerogel system, but this must be considered only as a crude estimation: we indeed recall that our treatment is not aimed at a quantitative description of actual systems because of the coarse-grained character of the model and the use of a (local) mean-field approximation.  
Note that the density of the adsorbed fluid is rescaled by  its value at $\mu=\mu_{sat}$ so that all the curves go to $1$ as $\mu\to \mu_{sat}$ (one has $\rho_f(\phi,\mu_{sat})\approx \phi\ \rho_l(T) $ where $\rho_l(T)$ is the density of the bulk liquid at coexistence).  

 \begin{figure}[hbt]
\includegraphics[width=8cm]{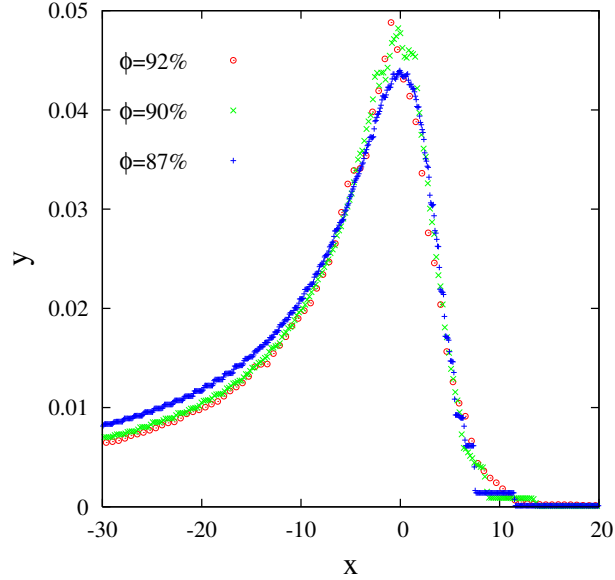}
  \caption{Scaling plot of the slope of the $T^*=0.4$ isotherms for $\phi=87\%, 90\%$ and $92\%$ according to Eq. 8 where $x=[\mu-\mu_{max}]/\mid \phi- \phi_c(T)\mid^{\beta \delta}$ and $y=\mid \phi- \phi_c(T)\mid^{\beta \delta-\beta}\rho_f(\phi,\mu_{sat})^{-1} d\rho_f(\phi,\mu)/d\mu $ with $\beta=0.1$, $\beta\delta=1.7$, and $\phi_c(0.4)=0.943$ (Color on line).} 
\end{figure}
Let us first study the scaling behavior when  varying the porosity  at fixed temperature. By analogy with the RFIM, our  basic scaling hypothesis is the existence, in the limit $L\to \infty$, of only two scaling variables $u(\phi,\mu)$ and $v(\phi,\mu)$ that measure the distance to the critical point  $(\phi_c(T), \mu_c(T))$.  Accordingly, the  adsorption isotherms in the vicinity of the critical point  $(\phi_c(T),\mu_c(T))$ should scale as
\begin{align}
\rho_f(\phi,\mu)-\rho_f(\phi_c(T),\mu_c(T))\sim \rho_f(\phi,\mu_{sat})\mid u/u_0\mid^{\beta} {\cal R}_{\pm}\Big (\frac{v/v_0}{ \mid u/u_0\mid^{\beta \delta}}\Big)
\end{align}
where ${\cal R}_{\pm}$ are the corresponding scaling functions and $u_0$ and $v_0$ are two nonuniversal scale factors that are temperature-dependent (they actually depend on all the specific features of the model\cite{PF1984}: $y$, the lattice structure, the range of interaction, etc...). 
Note that the scaling variables $u$ and $v$ need not be exactly $\phi- \phi_c(T)$ and $\mu-\mu_c(T)$, but their dependence on  $\phi$ and $\mu$ should be analytic\cite{C1996}. Unless one uses the proper variables, the curves do not collapse. In the case of the RFIM\cite{PDS1996}, it was shown that the proper scaling must be done with replacing $h=H-H_c$ in Eq. (6) by the ``rotated"  variable $h'=h+br$, where $b$ is some non-universal constant (on the other hand, there is no need to rotate $r$ because the corresponding correction is irrelevant when $\sigma \to \sigma_c$).  In practice, the rotation of the x-axis can be avoided by simply lining up the peaks in the derivative of the curves, with no correction terms, as is often done in analyzing experimental data (by considering  the slope $d\rho_f(\phi,\mu)/d\mu$ instead of $\rho_f(\phi,\mu)$ we also get  rid of the dependence on  $\rho_f(\phi_c(T),\mu_c(T))$). What to do on the y-axis is a matter of choice and, for simplicity, we just take $u=\phi- \phi_c(T)$. (Note that in Ref.\cite{PDS1996} the curves $dM/dH$ in the RFIM are plotted using $r=(\sigma-\sigma_c)/\sigma$ instead of the more standard  choice $r=(\sigma-\sigma_c)/\sigma_c$; alternatively, it has been proposed to expand $r$ to second-order in $(\sigma-\sigma_c)/\sigma_c $\cite{PV2003}; these different choices are equivalent asymptotically if no rotation is required, but they may change significantly the amplitude of the critical regions.) Therefore, to summarize, we look for a scaling of the form
 \begin{align}
\frac{1}{\rho_f(\phi,\mu_{sat})}\frac{d\rho_f(\phi,\mu)}{d\mu}\sim  a(T)\mid u\mid^{\beta-\beta \delta} \dot{{\cal R}}_{\pm}\Big(b(T)\frac{\mu-\mu_{max}}{\mid u\mid^{\beta \delta}}\Big)
\end{align} 
where $u=\phi- \phi_c(T)$, $\mu _{max}(\phi,T)$ is the value of the chemical potential at the maximum of $d\rho_f(\phi,\mu)/d\mu$, $\dot{{\cal R}}_{\pm}$ is the derivative of  ${\cal R}_{\pm}$, and $a(T)$ and $b(T)$ account for the dependence on $u_0$ and $v_0$.
Since it is important to have smooth isotherms so that one can take the derivative numerically, the average must be performed over a significant number of samples: the curves shown in Fig. 10 result from an average  over $500$ gel realizations of size $L=100$ (however, there are still some annoying irregularities in the average slopes, especially around $\mu=\mu_{max}$).

Calculations are of course done in finite systems so that one should also take into account finite-size effects\cite{PF1984}. These are especially important in the low-disorder regime ($\phi>\phi_c(T))$ in which there is a discontinuity in the isotherms in the thermodynamic limit. This introduces an additional difficulty in the analysis and it is a very demanding numerical task to perform a systematic finite-size scaling study of all the curves (see Ref.\cite{DKRT2003} for such a study of the $T^*=0.45$ isotherm in the  $95\%$ porosity aerogel). As a consequence, we shall only consider here the isotherms that are clearly in the strong disorder regime, which leads to reasonably size-independent results: this  eliminates  in particular the $T^*=0.4, 0.5$ and $0.6$  isotherms in the  $95\%$ porosity aerogel.

 Fig. 11 shows the scaling plot of the slope of the $T^*=0.4$ isotherms for $\phi=87\%,90\%$ and $92\%$ obtained with  the choice $\beta \delta=1.7$,  $\beta=0.1$, and $\phi_c(0.4)=0.943$ (the scale factors $a(T^*=0.4)$ and $b(T^*=0.4)$ are arbitrarily fixed equal to $1$).  It is actually very difficult to adjust independently the three unknown parameters  $\beta \delta$, $\beta$ and $\phi_c(T)$;  the values of the exponents  used here are the ones that allow for a collapse of a whole set of isotherms, as discussed below. It can be seen that the $\phi=92\%$ and $\phi=90\%$ isotherms collapse onto a single curve, which is not the case of the  $\phi=87\%$ isotherm,  especially on the left-hand  side of the peak. This part of the isotherm is indeed dominated by the presence of a liquid layer that coats the aerogel strands\cite{DKRT2003,DKRT2004},  and in the low-porosity gels the remaining available void space  is too small for allowing large avalanches to spread out. This  contribution to the adsorption has nothing to do with a critical phenomenon and it is reasonable to eliminate from  the scaling plot  all the isotherms obtained in the $87\%$ porosity aerogel. More generally, at each temperature, we choose to discard the isotherms that do not  belong to the scaling region because the porosity is obviously too far from $\phi_c(T)$ (i.e., the distance from the effective critical disorder is too large).  This finally leads to the global scaling plot shown in Fig. 12, obtained by adjusting also the nonuniversal scale factors  $a(T)$ and $b(T)$ (which, however, always remain of order $1$). 
 
The overall quality of the collapse is not perfect, but similar to that found in the RFIM\cite{PDS1996} (it is actually better because we only keep two isotherms at each temperature\cite{note11}). In fact, as discussed  in Ref.\cite{PDS1996}, the magnetization curves (or, here, the isotherms) are quantities that do not collapse  well, which makes it difficult to obtain from the scaling plots a precise determination of the critical exponents or of the location of the critical point. Although our estimated  values of the exponents $\beta \delta$ and $\beta$  are compatible with those of the RFIM,  the uncertainty is much too large to conclude that our model belongs to the same universality class. On the other hand, it is noteworthy that the value $1.7$ for the exponent  $\beta \delta$ is consistent with the value $1.22$ for the finite-size scaling exponent $\chi=\beta\delta/ \nu$ obtained in Ref.\cite{DKRT2003}, if one uses for $1/\nu$ the RFIM value, $0.71 \pm 0.09$\cite{PDS1996}.

\begin{figure}[hbt]
\includegraphics*[width=8cm]{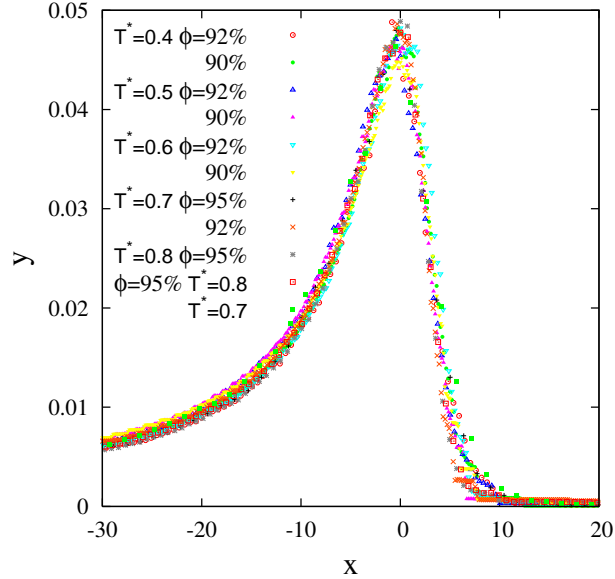}
  \caption{Scaling plot of the slope of the isotherms according to Eq. 8 (when varying the gel porosity at different temperatures)  or to Eq. 9 (when varying the temperature for the porosity $\phi=95\%$ ). $x$ and $y$  are the corresponding scaling variables and scaling functions (Color on line).}
  \label{fig:panorama}
\end{figure}

\begin{figure}[hbt]
\includegraphics*[width=9cm]{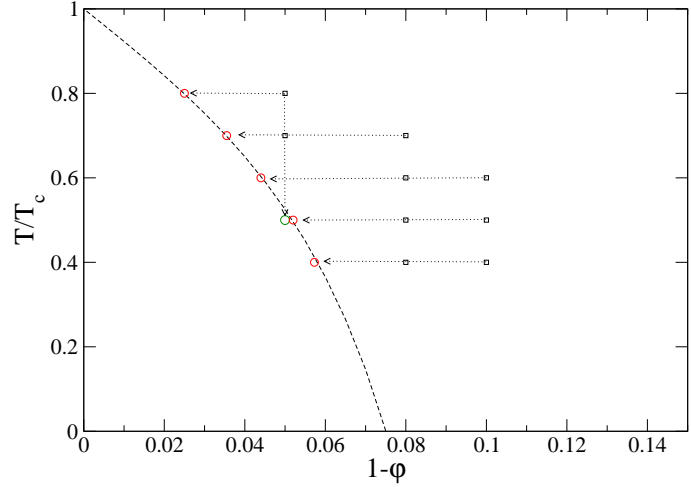}
  \caption{Approximate phase diagram of the model in the temperature-porosity plane deduced from the scaling plots. The dashed curve  is the critical line $T_c(\phi)$ (or $\phi_c(T)$) below which there is a first-order nonequilibrium transition in the adsorption isotherm. The squares indicate the isotherms that are included in the scaling plot of Fig. 12 (Color on line).}
  \label{}
\end{figure}

\begin{figure}[hbt]
\includegraphics*[width=8cm]{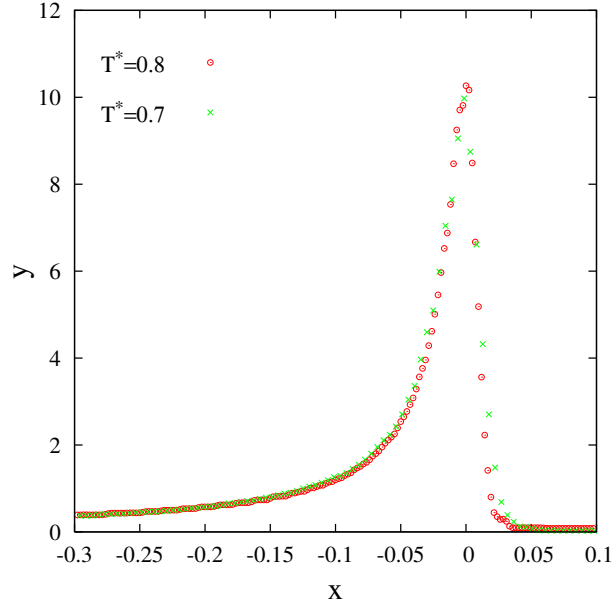}
  \caption{Scaling plot of the slope of the $T^*=0.8$ and $T^*=0.7$  isotherms in the $95\%$ aerogel according to Eq. 9 where $x=[\mu-\mu_{max}]/\mid T/T_c(\phi)-1\mid^{\beta \delta}$ and $y=\mid T/T_c(\phi)-1\mid^{\beta \delta-\beta} \rho_f(\phi,\mu_{sat})^{-1}d\rho_f(\phi,\mu)/d\mu  $ with $\beta=0.1$, $\beta\delta=1.7$, and $T^*_c(\phi=0.95)=0.5$ (Color on line).}
\end{figure}
From the values of $\phi_c(T)$ extracted from the scaling plots at each temperature, we can map out an approximate phase diagram of the model in the temperature-porosity plane.  This is shown in Fig. 13, where the critical line has been forced to extrapolate towards the zero-disorder limit $\phi_c(T_c)=1$. (Let us recall however that, in a real system, there cannot be a clear-cut transition close to $T_c$ because thermal fluctuations become larger and can allow the system to cross the free-energy barriers on the experimental time-scale; moreover, as noted before, we expect the fractal character of the microstructure in  very light aerogels to come into play.)  The present calculations indicate that the critical temperature drops very quickly with the porosity, and the extrapolation to $T=0$ predicts that there is no transition at finite temperature in gels of porosity smaller than $\phi_c(T=0)\approx 92.5\%$. This is in agreement with finite-size scaling calculations performed in the $87\%$ porosity aerogel that show that the isotherms are continuous at all temperatures in the thermodynamic limit. On the other hand, in the  $95\%$ porosity aerogel, the study of Ref.\cite{DKRT2003} predicts a transition at $T^*=0.45$, which is indeed below the critical line of Fig. 13. Morever, as shown in the preceding section, the avalanche size-distribution has a power-law behavior at $T^*=0.5$, which is the critical temperature extracted from the scaling plot ($T_c^*(\phi=95\%)=0.5\pm 0.02$). It turns out that the same critical temperature is also obtained by studying the scaling behavior when varying the temperature at fixed porosity. The scaling is now performed according to the expression 

\begin{align}
\frac{1}{\rho_f(\phi,\mu_{sat})}\frac{d\rho_f(\phi,\mu)}{d\mu}\sim  a'(\phi)\mid u\mid^{\beta-\beta \delta} \dot{{\cal R}}_{\pm}\Big(b'(\phi)\frac{\mu-\mu_{max}}{\mid u\mid^{\beta \delta}}\Big)
\end{align} 
where $u=(T-T_c(\phi))/T_c(\phi) $ and $a'(\phi)$ and $b'(\phi)$ are porosity-dependent scale factors. It is clear from Fig. 13 that this procedure can only be used with the $95\%$ porosity aerogel.
The scaling plot of the $T^*=0.8$ and $T^*=0.7$ isotherms  is shown in Fig. 14 (again, the $T^*=0.6$ isotherm is discarded because of the presence of important finite-size effects). It can be seen that a good scaling is obtained with the same values of the critical exponents as used before. Remarkably, the resulting scaling function is also the same and it can be included  in Fig. 12 by adjusting  the scale factors $a'(\phi)$ and $b'(\phi)$. This is a clear indication  that the critical behavior of the system is controlled by the same fixed point for all temperatures and porosities.

We now discuss the comparison of all these results with experiments.

\section{Discussion}

In this work and our earlier work\cite{DKRT2003,DKRT2004},  we have presented a detailed numerical study of the adsorption of $^4$He in silica aerogel that explains the modification in the shape of the isotherms as the temperature or the porosity is changed. We have found that the adsorption occurs via a sequence of irreversible condensation events (avalanches) of varying size and shape. The traditional picture of capillary condensation in mesoporous solids in which pores of increasing size fill up successively as the chemical potential or the gas pressure increases does not hold in aerogel, except, approximately, for the special category of spherical avalanches that occur at the end of the adsorption process.

Our calculations predict that there exists a line of critical points in the temperature-porosity diagram separating two regimes of continuous and discontinuous adsorption and characterized by a power-law behavior of the avalanche size distribution. There is a rather large scaling region where adsorption isotherms at different temperatures and  different porosities collapse onto a single universal curve. Although the system sizes considered in this work are too small to reach a definite conclusion, the estimated values of the critical exponents seem to be consistent with those of the three-dimensional field-driven RFIM at $T=0$\cite{PDS1996,PV2003}. 

The qualitative agreement between the simulated and experimental loops\cite{TYC1999,BH2003, LGPW2004} is a strong argument in favor of the validity of the above scenario. However, a complete validation still awaits the experimental description of the scaling behavior of the isotherms and the observation of a power-law behavior in the avalanche size distribution. This latter issue requires some additional comments. It may be indeed distressing that avalanches have not  been observed so far with  $^4$He in aerogel. In fact, the only exemple of avalanche behavior in adsorption/desorption experiments concerns the draining of  superfluid $^4$He  from Nuclepore\cite{LWH1996}, a porous material that contains a random spatial distribution of interconnected cylindrical pores. In this case, superfluidity appears to be a critical factor for the existence and propagation of avalanches via the presence of a superfluid film on the Nuclepore substrate that provides a coupling mechanism even between distant pores. It  is likely  however that such a mechanism is not necessary for avalanches to occur in aerogels because of very open character of the microstructure. 

There are actually several requirements for observing avalanches in sorption experiments, which are all satisfied in the case of the experiments with Nuclepore\cite{LWH1996}, but not in the case of the existing aerogel studies\cite{WC1990,TYC1999,BH2003, LGPW2004}. Let us first recall that, by definition, avalanches correspond to some collective reorganization of the system {\it in response} to a variation of the external driving force. If  one is changing the density rather than the chemical potential or the pressure (by introducing a controlled amount of gas into the sample and then measuring the pressure change), the system may follow a different path among the metastable states, and the vertical jumps in the isotherms that are the signature of the avalanches may not be observed\cite{note30} (this could explain that there are many data points along the adsorption isotherm at $T=2.34K$  in $98\%$ aerogel displayed in Ref.\cite{TYC1999}). Secondly, as already noted,  the observation time must be larger than the time the system takes to respond to the change in the external control variable. This can be done  by  continuously changing the chemical potential at a very slow rate, and measuring the amount of gas in the solid  regularly (every $x$ seconds).  In practice, this  procedure is only possible if the (local) equilibration time is short (for instance, it is smaller than one second with superfluid helium in Nuclepore). Unfortunately, the relaxation times in aerogel are much longer (of the order of two hours with $^4$He at low  temperature in the experiments  of Ref.\cite{BH2003}), and one has to change $\mu$ or $P$ by  a finite step and keep it fixed until  a new local equilibrium (i.e., a new metastable state) is reached.  With this procedure, the issue of having a  system with a long equilibration time boils down to stability of the experimental conditions and of course patience; in principle, one could then observe avalanches.  Yet, if $\Delta \mu$ or $\Delta P$ is not small enough, the avalanches may overlap and the resulting isotherm may appear  continuous. This is probably the situation  with the experiments in Ref.\cite{BH2003} where $\Delta P/P \sim 10^{-4}$ (to be compared with $\Delta P/P =\Delta \mu/(k_B T)\sim 10^{-7}$ in the Nuclepore experiments\cite{LWH1996}). 
Finally, a last requirement is that the sensitivity of the measurements is large enough to discriminate the jumps in the adsorbed fluid density. This condition is more easily satisfied when these jumps are large enough, which implies that one must be close to the critical line displayed in Fig. 13. Although one must be very cautious in comparing our temperature scale with the true experimental one, it seems that this last condition is also not satisfied in the existing experiments.

\acknowledgments
We thanks N. Mulders, E. Vives, and P. E. Wolf for  useful discussions.
The Laboratoire de Physique Th\'eorique de la Mati\`ere Condens\'ee is the UMR 7600 of
the CNRS.


\begin{thebibliography}{10}
\bibitem{note2} More precisely, we expect the number of solutions to scale exponentially with the system size.
\bibitem{note6} It is easy to see from Eq. (1) that the elements of the Hessian matrix ${\bf H}$ are given by $H_{ij}=k_BT \delta_{ij}\eta_i/[\rho_i(\eta_i-\rho_i)] -w_{ff}n_{ij}$ where $\delta_{ij}$ is the Kronecker symbol and $n_{ij}=1$ when $i$ and $j$ are nearest neighbors and zero otherwise. ${\bf H}$ can thus be diagonalized by a unitary transformation.
\bibitem{note20} The  threshold $\delta \rho_i>10^{-3}$ ensures that the avalanche size is computed with enough precision.
\bibitem{note7}  The calculation of $R_g$ is restricted to avalanches that do not span the whole lattice.
\bibitem{note12} We measure the asphericity via the parameter $A=[(R_{g,x}^2-R_{g,y}^2)^2+(R_{g,y}^2-R_{g,z}^2)^2+(R_{g,z}^2-R_{g,x}^2)^2]/2R_g^2$.
The normalisation  is such that $A=0$ for spherically symmetric objects and $1$ for ``cigar" shaped objects. Note that $A$ is not an invariant quantity with this definition since the components of the radius of gyration tensor are not taken along the principal axes (see J. Rudnick and G. Gaspari,  J. Phys. A. {\bf 19}, L191 (1986)). It is likely that, on average, we underestimate the actual deviation from sphericity.
\bibitem{note13} Note however that $\Delta \Omega$ in Eq. 5 is a small quantity compared to the bulk and surface contributions.
\bibitem{note11} except at $T^*=0.8$ where we only take into account the isotherm in the $\phi=95\%$ porosity aerogel..
\end{thebibliography}
\end{document}